\documentstyle [eqsecnum,aps] {revtex}
\tighten
\draft
\widetext
\input epsf
\newcommand{\beq}{\begin{equation}}
\newcommand{\eeq}{\end{equation}}
\newcommand{\beqs}{\begin{eqnarray}}
\newcommand{\eeqs}{\end{eqnarray}}
\newcommand{\lsim}{\mathrel{\raisebox{-
.6ex}{$\stackrel{\textstyle<}{\sim}$}}}
\newcommand{\gsim}{\mathrel{\raisebox{-
.6ex}{$\stackrel{\textstyle>}{\sim}$}}}

\begin{document}

\draft

\baselineskip 6.0mm

\title{Flavor-Changing Processes in Extended Technicolor 
}

\vspace{6mm}

\author{Thomas Appelquist$^{a}$ \thanks{email:
thomas.appelquist@yale.edu} \and
Neil Christensen$^{b}$ \thanks{email: neil.christensen@sunysb.edu} \and
Maurizio Piai$^{a}$ \thanks{email: maurizio.piai@yale.edu} \and
Robert Shrock$^{b}$ \thanks{email: robert.shrock@sunysb.edu}}

\vspace{6mm}

\address{(a) \ Physics Department, Sloane Laboratory \\
Yale University \\
New Haven, CT 06520}

\address{(b) \ C. N. Yang Institute for Theoretical Physics \\
State University of New York \\
Stony Brook, N. Y. 11794 }

\maketitle

\vspace{10mm}

\begin{abstract}

We analyze constraints on a class of extended technicolor (ETC) models from
neutral flavor-changing processes induced by (dimension-six) four-fermion
operators. The ETC gauge group is taken to commute with the standard-model
gauge group. The models in the class are distinguished by how the left- and
right-handed $(L,R)$ components of the quarks and charged leptons transform
under the ETC group. We consider $K^{0} - \bar K^0$ and other pseudoscalar
meson mixings, and conclude that they are adequately suppressed if the $L$ and
$R$ components of the relevant quarks are assigned to the same (fundamental or
conjugate-fundamental) representation of the ETC group. Models in which the $L$
and $R$ components of the down-type quarks are assigned to relatively conjugate
representations, while they can lead to realistic CKM mixing and intra-family
mass splittings, do not adequately suppress these mixing processes. We identify
an approximate global symmetry that elucidates these behavioral differences and
can be used to analyze other possible representation
assignments. Flavor-changing decays, involving quarks and/or leptons, are
adequately suppressed for any ETC-representation assignment of the $L$ and $R$
components of the quarks, as well as the leptons.  We draw lessons for future
ETC model building.
\end{abstract}

\pacs{14.60.PQ, 12.60.Nz, 14.60.St}

\vspace{16mm}

\newpage
\pagestyle{plain}
\pagenumbering{arabic}

\section{Introduction}

It is possible that electroweak symmetry breaks via the formation
of a bilinear condensate of fermions with a new strong
gauge interaction, generically called technicolor (TC).  
To communicate this symmetry breaking to the standard model
(technisinglet) fermions, one embeds  technicolor  in a
larger, extended technicolor (ETC) theory~\cite{etc}. To satisfy
constraints from flavor-changing neutral-current (FCNC) processes, the
ETC vector bosons that mediate generation-changing transitions
must have large masses. To produce the hierarchy in the masses of
the observed three generations (families) of fermions, the ETC
masses arise from the sequential breaking of the ETC gauge
symmetry on mass scales ranging from $10^3$ TeV down to the TeV
level. Precision measurements place tight constraints on these
models, suggesting that there are a small number of new degrees of
freedom at the TeV scale and that the technicolor theory has an
approximately conformal (``walking'') behavior with large anomalous
dimensions~\cite{wtc} -\cite{precision}.

A class~\cite{at94} -\cite{ckm} of ultraviolet(UV)-complete ETC models takes
the ETC dynamics to consist of a strongly interacting gauge theory whose gauge
group commutes with the standard model (SM) gauge group.  With the ETC
representation assignments of the SM fermions depending on their assignments
under the standard model, features such as intra-family mass splitting and CKM
mixing can emerge.  The ingredients to drive the ETC breaking are present. The
models are distinguished by how the left ($L$)- and right ($R$)-handed quarks
transform under the ETC group. This assignment must also be made for the
charged leptons, but the choice is not critical for the considerations of this
paper. The models include a mass-generation mechanism for neutrinos
\cite{nt,ckm}, although neutrino masses and mixing will not play an important
role here.

Here we analyze the consequences of (dimension-6) four-fermion operators that
occur in the effective theory at energies below $\Lambda_{TC}$ in this class of
ETC models, taking into account the multi-scale nature of the ETC gauge
symmetry breaking along with mixing between ETC interaction eigenstates to form
mass eigenstates.  For a discussion of other phenomenologically relevant
quantities, affected by dimension-5 operators, we refer the reader
to~\cite{dml,qdml}.

In Section~\ref{fcnc}, before specifying the details of the class of models of
interest here, we present an effective field theory argument leading to the
conclusion that ETC theories, even with walking, may generate
phenomenologically unacceptable flavor-changing neutral current (FCNC)
transitions, and describe a simple symmetry requirement for the underlying ETC
dynamics such as to eliminate this problem.  In Section~\ref{ourETC}, we
describe the structure of our class of UV-complete models.  In the subsequent
sections, we present estimates of the contributions of four-fermion operators
produced by ETC to $K^0 - \bar K^0$ mixing and other pseudoscalar mixing, as
well as other processes.  We show that when quarks of a given electric charge
couple vectorially (in the fundamental or anti-fundamental representation) to
the ETC gauge field, constraints from flavor-changing neutral current processes
can be satisfied. We also consider FCNC constraints when the $L$ and $R$ quarks
of a given charge are placed in relatively conjugate (fundamental and
anti-fundamental) ETC representations. If this is done for the down-type quarks
with the up-type quarks transforming vectorially, the model is capable of
producing adequate intra-family fermion mass splittings and
Cabibbo-Kobayashi-Maskawa (CKM) mixing. But this assignment does not suppress
FCNC processes sufficiently. In Section VI, we summarize and draw conclusions
for future model building, suggesting the use of other assignments.

\section{A generational symmetry and four-fermion operators.}
\label{fcnc}

ETC models, even in the presence of TC walking, can face a conflict between the
requirement of generating large enough masses for the standard model 
fermions and the simultaneous requirement of not generating unacceptably large
FCNC processes. In this section, we review
why this is the case. In the next section, we show that some models in the
class being considered can have the necessary symmetry structure to evade these
arguments.

Consider the effective theory at an energy $E>\Lambda_{TC}$ (the TC confinement
scale), but below all the (larger) ETC scales. The fermion spectrum consists of
all the SM fields and the technifermions. Having integrated out the ETC bosons,
the resultant four-fermion operators (all of which preserve the full SM  
and TC gauge symmetries) can be classified in three groups: those
involving only TC fields, which have no direct effect on the low-energy
phenomenology, so that we will not discuss them further, those involving only
ordinary fermions, and those that couple two technifermions and two
ordinary fields.  Each operator arises multiplied by the inverse square of an
ETC scale and a dimensionless coefficient.

To construct the effective theory for $E<\Lambda_{TC}$, physics at the TC-scale
is integrated out. The operators that couple
two technifermions and two ordinary fermions produce, through the formation of
TC-condensates, the dimension-3 bilinear fermion operators giving the ordinary
fermion mass matrices.
As an example, consider the mass of the down quark.  It arises from the
operator $[\bar d_L \gamma_\mu D_{L}][\bar D_{R} \gamma^\mu d_{R}]$, where $D$
is a down-type techniquark, and can be estimated to be
\beqs
m_d & \simeq & \kappa \eta \frac{\Lambda_{TC}^3}{\Lambda_{1}^2} \
,
\label{md11}
\eeqs
where $\Lambda_{1}$ is the highest ETC scale, associated with the
first family, $\kappa \sim O(10)$ is a numerical factor calculated
in Ref. \cite{ckm}, and $\eta$ is a factor incorporating walking,
which can plausibly be of order the ratio of the lowest ETC scale
to the TC scale ($O(10)$), but is unlikely to be  larger. A
realistic value for the mass $m_d$ can then be obtained naturally
with the (large) value $\Lambda_{1}\sim 10^3$ TeV, and with
$\Lambda_{TC} \sim 300$ GeV as dictated by the scale of
electroweak symmetry breaking.

The operators containing only ordinary
(technisinglet) fermions remain in the lower energy theory and is responsible
for the FCNC transitions of concern in this paper. 
Consider for example  
the $K^0\leftrightarrow \bar{K}^0$ mixing amplitude, 
generated by four-fermion operators of the form
$[\bar d_\chi \gamma_{\mu}s_\chi][\bar d_{\chi'}\gamma^{\mu}s_{\chi'}]$ ,
where $\chi\chi' = LL, LR, RR$. The standard model produces an operator of
this type with $\chi\chi'=LL$ and a coefficient that can fit experiment.  In
ETC models there are typically additional contributions to these operators. For
example, in certain ETC models there is a contribution to a $K^0 - \bar K^0$
operator (of LR type) at the scale $\Lambda_1$. Taking the coefficient of this
operator to be $b/\Lambda_1^2$, where $b$ is a
dimensionless number, the requirement of having a
sufficiently small contribution to $\Delta m_K$ implies
\beq
\frac{b}{\Lambda_1^2} \lsim \frac{{\rm few} \times 10^{-14}}
{{\rm  GeV}^2} \ .
\label{cdslim}
\eeq
With $b = {\cal O}(1)$, this would require $\Lambda_{1} \gsim 10^4$ TeV, an
order of magnitude larger than the (already large) value required to give a
realistic down-quark mass.

There is a natural way out of this problem, incorporated into some of the
models considered here. The four-fermion operators relevant for the down-quark
mass and for $K^0\leftrightarrow \bar{K}^0$ mixing, although generated at the
same scales, can have different symmetry properties with respect to the
underlying ETC gauge theory. This in turn can lead to $b \ll 1$ while the
corresponding coefficient in the mass-generating four-fermion operator is
$O(1)$. This is expected since the first operator involves four fields carrying
ordinary flavor quantum numbers, while the second involves just two.

For all the models we consider, the theory at energies below $\Lambda_{TC}$
contains an approximate global generational $U(1) ^3$ symmetry, with one $U(1)$
factor associated with each family of SM-fermions, and with each (chiral)
member of a family carrying its own $U(1)$ charge. This symmetry is a remnant
of the underlying ETC gauge symmetry, and the charge assignments are determined
by how each fermion transforms under the ETC gauge group. For the
suppression~(\ref{cdslim}) to be present, the left-handed and right-handed
down-type quarks of the first and second generation must be in ETC
representations such that the operator
$[\bar{d_L}\gamma_{\mu}s_L][\bar{d_R}\gamma^{\mu}s_R]$ violates the global
$U(1)^3$ symmetry. (Clearly, the corresponding operators with all four fields
of the same chirality violate the symmetry.)

If this is the case, then the $LR$ operator could not be generated by ETC
exchange if the global generational $U(1)^3$ symmetry were exact.  But in the
spontaneous breaking of the ETC gauge group, this symmetry is broken, by mixing
terms between different ETC gauge bosons. However, the mixing involves ratios
of the hierarchical scales of ETC symmetry breaking, and is therefore small,
strongly suppressing the contributions to the $LR$ operator.

\section{ETC Models}
\label{ourETC}

   We take the ETC gauge group $G_{ETC}$ to commute with the SM group $G_{SM} =
{\rm SU}(3)_c \times {\rm SU}(2)_L \times {\rm U}(1)_Y$.  The ETC group gauges
the three generations of technisinglet fermions and connects them with the
technicolored fermions.  We use $G_{ETC}={\rm SU}(N_{ETC})$, with the TC group
SU($N_{TC}) \subset G_{ETC}$.  For several reasons (see below), we choose
$N_{TC}=2$, and hence $G_{ETC}={\rm SU}(5)_{ETC}$.  The ETC gauge symmetry is
chiral, so that when it becomes strong, sequential breaking occurs naturally.
This breaking also involves one additional strongly coupled gauge interaction.
The breaking of the SU(5)$_{ETC}$ to SU(2)$_{TC}$ is driven by the
condensation of SM-singlet fermions which are part of the models; while there
is some freedom in the actual choice of these singlets, their presence is
always mandatory in order to cancel $SU(5)_{ETC}$ anomalies.  The SM-singlet
fermions that condense acquire large masses, and hence decouple from the
low-energy effective theory.

The ETC symmetry breaking takes place in stages, so that SU(5)$_{ETC} \to$
SU(4)$_{ETC}$ at a scale $\Lambda_1$, with the first-generation SM fermions
separating from the others; then SU(4)$_{ETC} \to$ SU(3)$_{ETC}$ at a lower
scale $\Lambda_2$ and SU(3)$_{ETC} \to$ SU(2)$_{TC}$ at a still lower scale
$\Lambda_3$, with the second- and third-generation fermions separating in the
same way, leaving the technifermions.  As ${\rm SU}(N)_{ETC}$ breaks to ${\rm
SU}(N-1)_{ETC}$ at the scale $\Lambda_j$, the $2N-1$ ETC gauge bosons in the
coset ${\rm SU}(N)/{\rm SU}(N-1)$ gain masses
\beq
M_j \simeq \frac{g_{_{ETC}} a \Lambda_j}{4}
\label{Mj}
\eeq
where $a \simeq O(1)$.  Since $g_{_{ETC}}^2/(4 \pi) \simeq O(1)$, it
follows that $M_j \simeq \Lambda_j$.  Following our earlier work
\cite{nt,ckm,dml,qdml}, we take these scales for definiteness to be
\beq
\Lambda_1 = 10^3 \ {\rm TeV}, \quad \Lambda_2 = 10^2 \ {\rm TeV},\quad
\Lambda_3 = 4 \ {\rm TeV} \ .
\label{lamscales}
\eeq
At the scale $\Lambda_{TC}$, technifermion condensates break the
electroweak symmetry.

The choice $N_{TC}=2$ has the advantages that it (a) minimizes the TC
contributions to the electroweak $S$ parameter, (b) with
a SM family of technifermions, $Q_L = {U \choose D}_L$,
$L_L = {N \choose E}_L$, $U_R$, $D_R$, $N_R$, $E_R$ transforming
according to the fundamental representation of SU(2)$_{TC}$, can yield
an approximate infrared fixed point and the associated walking
behavior \cite{wtc,nf} and (c) makes possible a mechanism to account for
light neutrinos without any super-heavy mass scale \cite{nt,lrs}.

Each of the above technifermions together with a set of ordinary fermions with
the same SM quantum numbers is placed in a representation of $SU(5)_{ETC}$. In
each case, the charge assignments of the components under the approximate
global $U(1)^3$ described in Section II depend on the $SU(5)_{ETC}$
representations.  For a fermion $(\psi_\chi)^{i_1...i_m}_{j_1...j_n}$ of
chirality $\chi=L,R$ transforming according to a general representation of
$SU(5)_{ETC}$, the charge $Q_k$ of a given component under the $k$'th $U(1)$ of
the $U(1)^3$, for $k=1,2,3$, is
\beq
Q_k \, = \,\sum_{p=1}^m \delta_{k,i_p} - \sum_{q=1}^n \delta_{k,j_q} .
\label{qcal}
\eeq
Consider, for example, the left-handed quark-techniquark electroweak
doublet. If it is assigned to the fundamental (anti-fundamental) ETC
representation, then the $U(1)^3$-charge assignment of its first-family
members is $(\pm1,0,0)$, etc. As we will see, more general representational
assignments may be necessary to produce fully realistic models.

In previous work we have analyzed two types of ETC models in the general class
\cite{at94} -\cite{qdml}.  In one (denoted CSM in Ref.  \cite{ckm}), $L$ quarks
and $R$ up-type quarks are assigned to the fundamental representation of
$SU(5)_{ETC}$, while the $R$ down-type quarks transform according to the
conjugate fundamental representation~\cite{ssvz}.  These models exhibit
charged-current CKM flavor mixing, intra-generational mass splittings without
excessive contributions to $\rho-1$ where $\rho=m_W^2/(m_Z^2 \cos^2\theta_W)$,
as well as the natural appearance of CP-violating phases. However, they give
rise to the operator $[\bar{d_L}\gamma_{\mu}s_L][\bar{d_R}\gamma^{\mu}s_R]$
without violating the $U(1)^3$ global symmetry, and therefore give an excessive
ETC contribution to the $K^0\leftrightarrow \bar{K}^0$ mixing amplitude.

In another type of model (denoted VSM in Ref. \cite{ckm}), the $L$ and $R$
components of all the quarks and techniquarks transform according to the
fundamental representation of the ETC group \cite{at94}-\cite{lrs}.  Without
additional ingredients these models cannot lead to realistic CKM mixing and
intra-family mass splittings. The vectorial structure of these models does,
however, naturally lead to adequate suppression of flavor-changing neutral
current processes (for example, the operator
$[\bar{d_L}\gamma_{\mu}s_L][\bar{d_R}\gamma^{\mu}s_R]$ in this case violates
the $U(1)^3$). Although the vectorial structure was typical of early ETC model
building, the natural FCNC suppression seems not to have been noticed.

Clearly neither type of model is fully realistic, and it will be important to
extend the class, exploring the assignment of the fermion fields to other
representations of the ETC group, and possibly including additional
interactions at energies not far above $\Lambda_1$. Such modifications are also
needed to eliminate one other problem with the class of models. They all have a
small number of unacceptable Nambu-Goldstone bosons arising from spontaneously
broken $U(1)$ global symmetries. These must be removed or given sufficiently
large masses to have escaped detection. In this paper, as we consider each
physical process, we take, for simplicity, the $L$ and $R$ components of the
relevant quarks (quarks of a given charge) to transform according to either the
same (fundamental or anti-fundamental) ETC representations or to relatively
conjugate (fundamental and anti-fundamental) representations. The latter
choice, as indicated above, will give excessive contributions to
$K^0\leftrightarrow \bar{K}^0$ and other mixing. We comment on other possible
ETC representation assignments in the summary Section VI.

\subsection{ETC Gauge Bosons}

Each SM quark or charged lepton of a given chirality $\chi=L,R$ is embedded in
a 5 or $\bar 5$ representation of SU(5)$_{ETC}$ so that the components with
indices $i=1,2,3$ correspond to the three generations, and those with indices
$i=4,5$ are the technifermions. For a fermion $f_\chi$ transforming as a 5 of
SU(5)$_{ETC}$, the basic coupling to the ETC gauge bosons is
\beq {\cal L} = g_{_{ETC}} \bar f_{j,\chi} (T_a)^j_k(V_a)^\lambda
\gamma_\lambda f^k_\chi \label{gff} \eeq
where the $T_a$, $a=1,...,24$ are the generators of the Lie
algebra of SU(5)$_{ETC}$ and the $V_a$ are the corresponding ETC
gauge fields. Similarly, if $f_\chi$ transforms as a $\bar 5$,
this coupling is $g_{_{ETC}} \bar f^j_{\chi} (T_a)^k_j
(V_a^\lambda) \gamma_\lambda f_{k \chi}$.  For nondiagonal
transitions, $j \ne k$, it is convenient to use the fields $V^j_k
= \sum_a V_{a,\lambda} (T_a)^j_k$, whose absorption by $f^k_\chi$
yields $f^j_\chi$, with coupling $g_{_{ETC}}/\sqrt{2}$, analogous
to the $W^\pm$ in SU(2)$_L$.  We take the diagonal (Cartan)
generators to be $T_{24} \equiv T_{d1}=(2\sqrt{10})^{-1}{\rm
diag}(-4,1,1,1,1)$, $T_{15} \equiv T_{d2}=(2\sqrt{6})^{-1}{\rm
diag}(0,-3,1,1,1)$, $T_8 \equiv T_{d3}=(2\sqrt{3})^{-1}{\rm
diag}(0,0,-2,1,1)$, and $T_3 = (1/2){\rm diag}(0,0,0,-1,1)$.  The
ETC gauge bosons that couple to these diagonal generators $T_{dj}$
are denoted $V_{dj}$.

When SU(5)$_{ETC}$ breaks to SU(4)$_{ETC}$, the nine ETC gauge bosons in the
coset SU(5)$_{ETC}$/SU(4)$_{ETC}$, namely, $V^1_j$, $(V^1_j)^\dagger=V^j_1$,
$j=2,3,4,5$, and $V_{d1}$, gain masses $M_1 \simeq \Lambda_1$. Similarly, when
SU(4)$_{ETC}$ breaks to SU(3)$_{ETC}$, the seven ETC gauge bosons $V^2_j$ and
$(V^2_j)^\dagger=V^j_2$, $j=3,4,5$, together with $V_{d2}$, gain masses $\simeq
\Lambda_2$.  Finally, when SU(3)$_{ETC}$ breaks to SU(2)$_{TC}$, the five ETC
gauge bosons $V^3_j$, $(V^3_j)^\dagger=V^j_3$, $j=4,5$, together with $V_{d3}$,
gain masses $\simeq \Lambda_3$. The SM-singlet fermions responsible for this
breaking also, through quantum loops, lead to mixing among these gauge bosons,
so that they are not exact mass eigenstates.  The mixing is small, being
suppressed by ratios of the hierarchical ETC scales.

These mixing terms among ETC gauge bosons are the source of breaking of the
global generational $U(1)^3$ symmetry introduced in Section~\ref{fcnc}. In the
absence of any such mixing, this symmetry would remain unbroken, and would
forbid many four-fermion operators in the effective theory for
$E<\Lambda_{TC}$.  The smallness of the ETC mixing, together with the largeness
of the ETC scales, will be crucial for the suppression of FCNC processes.

A particular type of ETC mixing will be focused on in this paper. This is the
mixing among the ETC gauge bosons $V_t^i$ that transform as doublets under
SU(2)$_{TC}$ (with ETC index $t \in \{4,5\}$) and triplets under the
generational SU(3) (with ETC index $i \in \{1,2,3\}$, $j \ne k$). This can be
of the form $V_t^{i} \leftrightarrow V_t^j$, producing the off-diagonal
elements of the quark mass matrices when the $L$ and $R$ components transform
according to the same representation. We note that the diagonal elements do not
require mixing since $f_L^i$ and $f^{i}_{R}$ have the same charge under the
$U(1)^3$ generational symmetry introduced in Section II. Thus the off-diagonal
elements are suppressed relative to the diagonal elements. The ETC mixing can
also be of the form $V_4^i \leftrightarrow V_j^5$ when the $L$ and $R$
components transform according to relatively conjugate ETC-representations.  In
this case, $f_L^i$ has a different $U(1)^3$-charge than $f_{j\,R}$ for all
($i,j$), and therefore all the elements of the mass matrix are suppressed.

\subsection{Quark Masses}

The effective theory describing the physics at energies $E<\Lambda_{TC}$,
obtained by integrating out the ETC and TC gauge bosons and all the heavy
fermions, contains the mass matrix of SM quarks or charged leptons, given by
\beq
{\cal L}_m = -\bar f_{j,L} M^{(f)}_{jk} f_{k,R} + h.c.,
\label{lm}
\eeq
where $f$ denotes up-type and down-type quarks, as well as charged
leptons, and the indices $j,k \in \{1,2,3\}$ are generation
indices, all written as subscripts here. 
The structure of $M^{(f)}$ depends on the type of ETC
model that generates it. If the $L$ and $R$ components transform
according to the same representation of the ETC group, then the
diagonal elements of $M^{(f)}$ do not require any ETC gauge boson
mixing (being invariant under $U(1)^3$),
while the off-diagonal elements do require mixing. If the
$L$ and $R$ components transform according to relatively conjugate
representations, then ETC mixing is required for all the elements
of $M^{(f)}$.

An arbitrary mass matrix $M^{(f)}$ can be brought to real, positive
diagonal form by the bi-unitary transformation
\beq
U^{(f)}_L M^{(f)} U^{(f) \ -1}_R = M^{(f)}_{diag.} \ .
\label{biunitary}
\eeq
Hence, the interaction eigenstates $f$ of the quarks are mapped to mass
eigenstates $q$ via
\beq
f_\chi = U^{(f) \ -1}_\chi q_{\chi}
\label{ffm}
\eeq
for $\chi=L,R$.  Writing out the vectors $q_\chi$ explicitly for
the up and down quarks, we have $u_\chi = (u,c,t,U^4,U^5)_\chi$,
$d_\chi=(d,s,b,D^4,D^5)_\chi$, and, for the leptons,
$(e,\mu,\tau,E^4,E^5)_\chi$.  Here we use the notation $u$ and $d$
to refer to the respective charge $Q=2/3$ and $Q=-1/3$
five-dimensional representations of SU(5)$_{ETC}$ and to the
individual $u$ and $d$ quarks; the specific meaning will be clear
from context.  The observed CKM quark
mixing matrix $V$ that enters in the charged weak current is then
given by
\beq
V=U^{(u)}_L U^{(d) \ \dagger}_L \ .
\label{vckm}
\eeq

For the parametrization of the matrices $U^{(f)}_\chi$, we recall
that a general $N \times N$ unitary matrix $U$ (where here $N =3$
generations) can be written as $U = e^{i\phi} {\cal U}$ where
${\cal U} \in {\rm SU}(N)$. The matrix $U$ depends on $N^2$ real
parameters, of which $N(N-1)/2=3$ are rotation angles, and the
remaining $N(N+1)/2=6$ are complex phases.  Thus each of the
matrices $U^{(f)}_\chi$, $\chi=L,R$, depends on three angles
$\theta^{(f)\chi}_{mn}$, $mn =12,13,23$, and six (independent)
phases.  Some of these phases can be removed by rephasings of
quark fields, as we discuss further below. We parametrize the
transformation matrices $U^{(f)}_\chi$, $\chi=L,R$, as
\beq
U^{(f)}_\chi = e^{i\phi_{(f)\chi}} \,P_\alpha^{(f) \chi}\, U_{0
\chi}^{(f)}\,
P_\beta^{(f) \chi}
\label{ufchi}
\eeq
where
\beq
P_\alpha^{(f) \chi}={\rm diag}(e^{i\alpha_1^{(f)\chi}},
e^{i\alpha_2^{(f)\chi}},e^{i\alpha_3^{(f)\chi}}) \ ,
\label{pachi}
\eeq
\beq
P_\beta^{(f) \chi}={\rm diag}(e^{i\beta_2^{(f)\chi}},
e^{i\beta_2^{(f)\chi}},e^{i\beta_3^{(f)\chi}}) \ ,
\label{pbchi}
\eeq
with $\alpha_3^{(f)\chi}=-\alpha_1^{(f)\chi}-\alpha_2^{(f)\chi}$,
$\beta_3^{(f)\chi}=-\beta_1^{(f)\chi}-\beta_2^{(f)\chi}$,
and the matrix $U_0^{(f)}$ follows the ordering conventions of
\cite{pdg},
\beq
U_{0 \chi}^{(f)} = R_{23}(\theta_{23}^{(f)\chi}) \, P^{(f)\chi \
*}_\delta \,
R_{13}(\theta_{13}^{(f)\chi}) \, P^{(f)\chi}_\delta \,
R_{12}(\theta_{12}^{(f)\chi}) \ ,
\label{uofchi}
\eeq
where $R_{mn}(\theta^{(f)\chi}_{mn})$ is the rotation in the $mn$
subsector,
and
\beq
P^{(f)\chi}_\delta={\rm diag}(e^{i\delta^{(f)\chi}},1,1) \ .
\label{pdelta}
\eeq
It is a convention \cite{pdg} how one chooses to insert a phase like $\delta$
among the rotations $R_{12}$, $R_{13}$, and $R_{23}$, but physical quantities
depend only on expressions that are independent of such conventions. Note that
we have not tried to remove a maximal number of phases to put the resultant
quark mixing matrix $V$ in its canonical form.  When dealing with CP-violating
quantities, we will therefore write explicitly rephasing-invariant expressions.

In the special case when the $L$ and $R$ components of the fields
$f$ transform in the same
ETC representation, the vectorial nature
of the ETC interactions responsible for generating
their mass matrix implies
\beq
U^{(f)}_L = e^{-i\,\phi_{(f) R}} U^{(f)}_R \equiv U^{(f)}\,,
\label{ulur}
\eeq
where $U^{(f)}$ is unimodular. In this case, $M^{(f)}$ is hermitian
up to the phase factor $e^{-i\,\phi_{(f) R}}$.
Consequently, $\phi_{(f) L} =0$, $U_0^{(f)L}=U_0^{(f)R} \equiv U_0^{(f)}$,
$P_\alpha^{(f)L}=P_\alpha^{(f)R} \equiv P_\alpha^{(f)}$, etc.  and thus
$\theta^{(f)L}_{mn}=\theta^{(f)R}_{mn} \equiv \theta^{(f)}_{mn}$,
$\delta^{(f)L}=\delta^{(f)R} \equiv \delta^{(f)}$, $\alpha^{(f)L}_j =
\alpha^{(f)R}_j \equiv \alpha^{(f)}_j$, and $\beta^{(f)L}_j = \beta^{(f)R}_j
\equiv \beta^{(f)}_j$.

Since we will analyze CP-violation in neutral $K$ and $B_d$ mixing, a remark
on the strong CP problem is in order.  This is the problem of why
$|\bar\theta| \lsim 10^{-10}$, where $\bar\theta = \theta
-[arg(det(M^{(u)})) + arg(det(M^{(d)}))]$, with $\theta$ appearing via the
topological term $\theta g_s^2(32 \pi^2)^{-1} G_{a \ \mu\nu} \tilde
G_a^{\mu\nu}$.  Whether a resolution of the strong CP problem will emerge in
the class of models considered here is not yet clear \cite{lane_kkb}.
However, whatever the resolution of the strong CP problem turns out to be,
$\bar\theta$ involves only the flavor-independent $\phi_{(f)\chi}$ phases in
the $U^{(f)}_\chi$, $\chi=L,R$. By contrast, the CP-violating quantities
considered here and in Ref. \cite{qdml} depend on the other,
generation-dependent phases in the $U^{(f)}_\chi$. Even in a theory that
provides for a solution of the strong CP problem, one must analyze the
effects of these flavor-dependent phases, as we do here.  Aside
from assuming that $|\bar\theta|$ is sufficiently small, we will not make
any special assumptions concerning the sizes of the CP-violating phases that
enter in the $U^{(f)}_\chi$.  This is in accord with the fact that the
intrinsic CP violating phase $\delta$ in the CKM matrix, is not small
\cite{parodi}.

\subsection{Dimension-6 Four-Fermion Operators}

Integrating out the ETC-scale physics produces not only the quark-mass
operators, but also operators of higher dimension. In previous papers, we
discussed the impact of the (dimension-5) dipole operators
\cite{dml,qdml}. Here we are interested in the (dimension-6) four-fermion
operators describing flavor-changing neutral-current processes. Since at each
scale of ETC breaking, the ETC interactions are strong, the estimate of the
four-fermion operators must be done to all orders in this coupling.

The four-fermion operators of interest receive two types of contributions from
the exchange of heavy ETC gauge bosons, with ETC mixing and without. If the
fermion assignments to the ETC gauge group are such that a four-fermion
operator preserves the $U(1)^3$ global symmetry, then its coefficient is
suppressed only by the mass scale of the gauge boson exchanged; no ETC mixing
is required. (For the processes considered in this paper, this scale will be
$\Lambda_1$, with the exception of $B_s - \bar B_s$ mixing, for which it will
be $\Lambda_2$.) This will be problematic for certain neutral-pseudoscalar
mixing processes.

If a four-fermion operator does not respect the $U(1)^3$ global symmetry,
its generation requires ETC mixing (which violates the global symmetry).
This mixing can take place, for example, among the ETC gauge bosons
transforming as $SU(2)_ {TC}$-doublets. It enters the four-fermion operators
through the unitary matrices that diagonalize the fermion mass matrices. The
coefficient of the resultant four-fermion operators is then suppressed not
only by the masses of the exchanged ETC gauge bosons, but also by (small)
mixing angles coming from Eqs.~(\ref{ffm})-(\ref{uofchi}).

The ETC mixing and associated breaking of the global $U(1)^3$ generational
symmetry takes place also among the ETC gauge bosons transforming as TC-
singlets. This mixing is not directly responsible for the generation of the
SM-fermion mass matrices, but appears among the ETC gauge bosons exchanged
between the SM-fermions. These mixings, too, are proportional to ratios of ETC
scales, and hence suppressed, so that the contribution to four-fermion
operators is  small.  There will be relations between this type of mixing
and that among the $SU(2)_{TC}$-doublet gauge bosons. But we expect its effect
on four-fermion operators to be at most comparable to that due to the doublet
mixing, and therefore we will not consider it further in this paper.

To estimate the contribution to the four-fermion operators arising from the
off-diagonal quark mass matrices, it will be helpful to re-express the ETC
gauge couplings in terms of quark mass eigenstates. The full $5 \times 5$
matrix $\sum_{a=1}^{24} T_a V_a^\lambda$ that enters this coupling, may be
restricted to its $3 \times 3$ submatrix involving only ordinary quark indices.
Suppressing the Lorentz index $\lambda$, we thus define
\beq
{ \cal V} = \left(\begin{array}{ccc}
-\frac{2V_{d1}}{\sqrt{10}}&\frac{V_2^1}{\sqrt{2}}&\frac{V_3^1}{\sqrt{2}}\\
\frac{V_1^2}{\sqrt{2}}&\frac{V_{d1}}{2\sqrt{10}}-\frac{3V_{d2}}
{2\sqrt{6}}&\frac{V_3^2}{\sqrt{2}}\\
\frac{V_1^3}{\sqrt{2}}&\frac{V_2^3}{\sqrt{2}}&\frac{V_{d1}}
{2\sqrt{10}}+\frac{V_{d2}}{2\sqrt{6}}-\frac{V_{d3}}{\sqrt{3}}
\end{array}\right)
\label{v}
\eeq

If both the $L$ and $R$ components are in the fundamental representation, the
coupling is vectorial and given by
\beq
{\cal L}_{int} = g_{_{ETC}} \sum_{f,j,k}
\bar f_j \gamma_\lambda ({\cal V}^\lambda)^j_k f^k
\equiv g_{_{ETC}} \sum_{q,j,k}
\bar q_j \gamma_\lambda (A^\lambda)^j_k  q^k \ ,
\label{coupling}
\eeq
with
\beq
A^{\lambda} \equiv U^{(f)} {\cal V}^\lambda U^{(f) \ -1} \ ,
\label{a}
\eeq
where we have used Eq. (\ref{ulur}), and we
recall from Eq. (\ref{ffm}) that the $f^j$ and $q^k$ are the ETC interaction
and mass eigenstates of the quarks of a given charge.  We can simplify this
expression by absorbing the $P^{(f)}_\alpha$ in the $q$ fields so that for
these ETC models with vectorial SM fermion representations,
\beq
A^\lambda = U_0^{(f)} P_\beta^{(f)} {\cal V}^\lambda P_\beta^{(f) \
*} U_0^{(f) \ \dagger} \ ,
\label{avsm}
\eeq
which does not depend on $P^{(f)}_\alpha$.  The above rephasing of the $q$
fields leaves the diagonalized fermion mass matrix invariant.

If the $L$ and $R$ components of the quarks transform according to
conjugate representations, then the ETC couplings are chiral and
given by
\beq
{\cal L}_{int,L} = g_{_{ETC}} \sum_{f,j,k,}
\bar f_{j,L} \gamma_\lambda ({\cal V}^\lambda)^j_k f^k_L
\equiv g_{_{ETC}} \sum_{f,j,k,\chi}
\bar q_{j,L} \gamma_\lambda (A^{(f) \ \lambda}_L)^j_k  q^k_L \ ,
\label{coupling_csm_L}
\eeq
and
\beq
{\cal L}_{int,R} = g_{_{ETC}} \sum_{f,j,k,}
\bar f^j_R \gamma_\lambda ({\cal V}^\lambda)^k_j f_{k,R}
\equiv g_{_{ETC}} \sum_{f,j,k,\chi}
\bar q^j_R \gamma_\lambda (A^{(f) \ \lambda}_R)^k_j  q_{k,R} \ ,
\label{coupling_csm_R}
\eeq
where
\beq
A^{(f) \ \lambda}_\chi = U^{(f)}_\chi {\cal V}^{\lambda} [U^{(f)}_\chi]^{-1} \ .
\label{achiral}
\eeq

\section{Neutral Pseudoscalar Meson Mixing}

Owing to the transitions $M^0 \leftrightarrow \bar M^0$, where $M^0 =K^0$,
$B_d$, $B_s$, or $D^0$, the mass eigenstates of these neutral
non-self-conjugate mesons involve linear combinations of $|M^0\rangle$ and
$|\bar M^0\rangle$.  The time evolution of the $M^0,\bar M^0$ system is
governed by  $M - i \Gamma/2$, where $M$ and $\Gamma$ are $2
\times 2$ hermitian matrices in the basis $(M^0,\bar M^0)$. The resultant
physical mass eigenstates have different masses; we denote these as $M^0_h$
and $M^0_\ell$, with mass difference $\Delta m_M = m_{M^0_h}-m_{M^0_\ell}$.
For the kaon system, $K_L = K_h$ and $K_S = K_\ell$ are the long- and
short-lived eigenstates, and $\Delta m_K = 2 {\rm Re} (M_{12})$.
For the $B$ and $D$ mesons, $M_h$ and $M_\ell$ have
essentially the same lifetimes, and hence
$\Delta m_{B,D} = 2 | M_{12} |$.  Direct experimental measurements and limits
on resultant mass differences are \cite{pdg,pitts}:
\beqs
\Delta m_K & = & (0.530 \pm 0.001) \times 10^{10} \ {\rm s}^{-1} =
(3.49 \pm 0.006) \times 10^{-15} \ {\rm GeV}\\
\Delta m_{B_d} & = & (0.502 \pm 0.007) \times 10^{12} \ {\rm s}^{-1} =
(3.36 \pm 0.04) \times 10^{-13} \ {\rm GeV}\\
\Delta m_{B_s} & > & 14.4 \times 10^{12} \ {\rm s}^{-1} = 0.99 \times
10^{-11} \ {\rm GeV} \quad (95 \ \% \ {\rm CL}) \\
\Delta m_{D} & < & 7 \times 10^{10} \ {\rm s}^{-1} =
0.5 \times 10^{-13} \ {\rm GeV}  \quad (95 \ \% \ {\rm CL})  .
\label{dm_values}
\eeqs

The standard model accounts for the two measured mass differences, $\Delta m_K$
and $\Delta m_{B_d}$ and agrees with the limits on the other two mass
differences $\Delta m_{B_s}$ and $\Delta m_D$ \cite{pdg,parodi}.  This thereby
constrains non-SM contributions such as those from ETC gauge boson exchanges.
For example, a recent SM fit gives $\Delta m_{B_s} \simeq (18-21) \times
10^{12}$ sec$^{-1}$ $=(1.2 - 1.4) \times 10^{-11}$ GeV with uncertainties $\sim
\pm 3 \times 10^{12}$ sec$^{-1}$, i.e. $\sim 0.2 \times 10^{-11}$ GeV
\cite{parodi}.  Evidently, this is rather close to the current lower limit
\cite{pdg}.  The standard model predicts that $\Delta m_D \sim
O(10^{-17})$ GeV \cite{pdg}, much smaller than its current experimental upper
limit.

We denote the effective Hamiltonian density for the transition $M^0
\leftrightarrow \bar M^0$, where $M^0 = q_j \bar q_k$, as
\beq
{\cal H}_{eff} =\sum_{\chi,\chi^\prime} c_{\bar j k; \chi\chi^\prime}
{\cal
O}_{\bar j k; \chi\chi^\prime}
\label{heff}
\eeq
with
\beq
{\cal O}_{\bar j k; \chi\chi^\prime} = [\bar q_{j \chi} \gamma_\lambda
q_{k \chi}] [\bar
q_{j \chi^\prime} \gamma^\lambda q_{k \chi^\prime}]
\label{operator_chiral}
\eeq
where $\chi$ and $\chi'$ denote chirality and the $c_{\bar j k;
\chi,\chi^\prime}$ are coefficients with dimensions of inverse mass squared.
Summations over color and spinor indices are understood. We then have
$M_{12} = \langle M^0 | \int d^3x {\cal H}_{eff} |\bar M^0 \rangle$

While SM box diagrams contribute only to ${\cal O}_{\bar j k; LL}$, ETC
gauge boson exchange can contribute to ${\cal O}_{\bar j k; \chi \chi'}$
with $\chi \chi'=LL,LR,RR$. Indeed, if the ETC couplings to the quarks are
vectorial, the ETC contribution to the effective Hamiltonian density for
$M^0 \leftrightarrow \bar M^0$ transitions has the simple form
\beq
{\cal H}_{eff,ETC} = c_{\bar j k} \, [\bar q_j \gamma_\lambda q_k][\bar
q_j \gamma^\lambda q_k] \ .
\label{operator_vectorial}
\eeq

Other operators are induced by renormalization group running. We neglect
these renormalization effects here, since they do not affect substantially
our results. We
calculate the ETC contributions to the coefficients of the above
four-fermion operators at the scale $\Lambda_{TC}$, studying their
dependence on the ETC-breaking scales and on the mixing angles in the quark
mass matrices.  To obtain physical predictions, we then sandwich the
operator in Eq. (\ref{operator_chiral}) between $M^0$ and $\bar M^0$ states,
taking account of the two different color contractions \cite{st}, and use,
as input, estimates of the relevant hadronic matrix elements $\langle M^0 |
{\cal O}_{\bar j k; \chi \chi'} |\bar M^0 \rangle$.  For recent lattice
measurements of these matrix elements, see, e.g., \cite{lattice}.

\subsection{$K^0 - \bar K^0$ Mixing}
\label{kk}

Let us start from  the (CP-conserving) mass difference
$\Delta m_K = 2 {\rm Re}(M_{12})$,
assuming that the $L$ and $R$ components of the down-type quarks
transform either according to the same (fundamental) ETC representation or
according to relatively conjugate representations.
The two cases
are very different, hence we treat them separately.

\subsubsection{Conjugate Representations}

To construct the amplitude $s \bar d \to d \bar s$ for the case of conjugate
representations, we need the $s-d$ coupling to ETC vector bosons expressed
in terms of fermion mass eigenstates given in Eqs. (\ref{coupling_csm_L})
and (\ref{coupling_csm_R}).  The key point is that there is a contribution
to the amplitude without the necessity of any ETC mixing \cite{ckm}. This is
due to the fact that with this assignment, $d_L$ and $s_L$ both have
$U(1)^3$ generational charges that are opposite to the charges of $d_R$ and
$s_R$, and hence $[\bar d_L \gamma_{\lambda} s_L][\bar d_R \gamma_{\lambda}
s_R]$ is invariant under the $U(1)^3$ generational symmetry.

Specifically, an $s_L \bar d_L$ pair in the initial-state $\bar K^0$ can
annihilate to produce a $V^2_1$. It can then directly create a $d_R \bar s_R$
in the final-state $K^0$ because the right-handed components transform
according to the conjugate fundamental representation. Similarly, a $s_R \bar
d_R$ pair in the initial $\bar K^0$ can produce a $V^1_2$ that directly creates
a $d_L \bar s_L$ in the final $K^0$.  Since ETC is strongly interacting at the
relevant scale, $\Lambda_1$, the lowest-order ETC amplitude provides only a
rough estimate, which is
\beq c_{\bar d s;LR} \simeq \left ( \frac{g_{_{ETC}}}{\sqrt{2}} \right )^2
\frac{\zeta}{M_1^2} \simeq \frac{8 \zeta}{\Lambda_1^2}
\label{KKb_CSM_tree} \eeq
where $\zeta$ is a phase factor of unit modulus, and we have used
Eq. (\ref{Mj}).  While higher-order ETC contributions are important, it is not
expected that they will substantially modify this estimate.  Even with
$\Lambda_1$ as large as $10^3$ TeV, as given in Eq. (\ref{lamscales}), the
estimate (\ref{KKb_CSM_tree}) leads to a value of $\Delta m_K$ that is nearly
two orders of magnitude larger than the experimental value \cite{ckm}.  As
emphasized already, we regard this as a serious problem for these models,
despite their success at producing intra-generational mass splittings.

It is possible for terms due to mixing to cancel the above contribution so as
to produce an acceptably small result, but this would require
that these two terms have similar magnitudes.  This could
happen, since contributions due to fermion mixing which are nominally
proportional to $1/\Lambda_j^2$ with $j=1$ or $j=2$ involves mixing angle
factors that are naturally small, so the actual size of such terms might be as
small as $1/\Lambda_1^2$.  However, we regard this as very unlikely, since
there is no symmetry reason for it and the hadronic matrix elements are
different; it would thus require fine-tuning.

\subsubsection{Vectorial Representation}

With both the $L$ and $R$ components of the down-type quarks in the fundamental
ETC representation, all the four-fermion operators entering the $ K^0 - \bar
K^0$ amplitude violate the global $U(1)^3$ generational symmetry. The $s
\bar d$ quark pair in the initial $\bar K^0$ has the generational quantum
numbers (i.e., ETC group index structure) given by $V^2_1$, and this cannot
directly (without ETC mixing) produce the $d \bar s$ quarks in the
final-state $K^0$ with its ETC group index structure given by $V^1_2$. In
order for this transition to proceed, ETC mixing is necessary. It must
transform the initial state with the ETC generational index structure of
$V^2_1$ to the final state with the structure of $V^1_2$.  This occurs via
loops of SM-singlet fermions at ETC scales below $\Lambda_1$ \cite{ckm} and
hence leads to strong suppression of the amplitude.

An example of the relevant ETC mixing is that among the SU(2)$_{TC}$-doublet
ETC bosons, leading to the off-diagonal quark mass matrices. It is
incorporated in the couplings (\ref{coupling}).  For this transition, we
need the quantity $(A^1_2)^\lambda$ appearing in the vertex $\bar d
\gamma_\lambda (A^1_2)^\lambda s$ in Eq. (\ref{coupling}). We keep mixing
terms involving couplings to the massive ETC vector bosons with the lowest
two masses, $\Lambda_3$ and $\Lambda_2$, and perform a Taylor series
expansion in small rotation angles, truncating it after the most relevant
terms:
\beqs
(A^\lambda)^1_2 & = & e^{i(\beta_1^{(d)}-\beta_2^{(d)})}
\frac{(V^\lambda)^1_2}{\sqrt{2}}
  - e^{-i\delta^{(d)})}
\theta_{13}^{(d)}\theta_{23}^{(d)}\frac{(V^\lambda)_{d3}}{\sqrt{3}}
- \theta_{12}^{(d)}\ \frac{3 (V^\lambda)_{d2}}{2\sqrt{6}} \cr\cr
& + & e^{i(\beta_3^{(d)}- \beta_2^{(d)}-\delta^{(d)})}\theta_{13}^{(d)}
\frac{(V^\lambda)^3_2}{\sqrt{2}}
+ e^{i(\beta_2^{(d)}- \beta_3^{(d)})}
\theta_{12}^{(d)}\theta_{23}^{(d)}\frac{(V^\lambda)^2_3}{\sqrt{2}} \ .
\label{a12}
\eeqs
Note that we have used the convention of Eq. (\ref{avsm}) in which the
$\alpha$ phases have been rotated away.

Consider the exchange of an ETC gauge boson between the quarks, in both $s$-
and $t$-channels. Wick-contracting the ETC gauge fields $(A^\lambda)^1_2$,
keeping only the leading small-rotation-angle terms and setting the momentum
in the ETC gauge boson propagator to zero, we have
\beq
c_{\bar d s} {\cal O}_{\bar d s} \simeq \biggl [
\frac{6}{\Lambda_2^2}(\theta_{12}^{(d)})^2 +
\frac{16}{3 \Lambda_3^2}e^{-2i\delta^{(d)}}
(\theta_{13}^{(d)}\theta_{23}^{(d)})^2 \biggr ]
[\bar d \gamma_\lambda s] [\bar d \gamma^\lambda s] \ .
\label{kkbar_v33}
\eeq
In this and subsequent expressions, we use the relation in Eq.  (\ref{Mj})
to re-express $g_{_{ETC}}^2/M_j^2 \simeq 16/\Lambda_j^2$ for $j=1,2,3$.
Higher-order ETC contributions are important because of the strong-coupling
nature of the ETC theory at this scale.  They are incorporated in the
coefficient of order unity that implicitly multiplies the right-hand side of
Eq. (\ref{kkbar_v33}).  We insert the operator (\ref{kkbar_v33}) between
$|\bar K^0 \rangle$ and $\langle K^0|$ states and perform the $\int d^3x$
integral to obtain $M_{12}$.

Since the standard model can fit the experimental value of $\Delta m_K$ 
up to the uncertainty due to long-distance QCD effects in the
calculation of this quantity, we require that the ETC contribution to
$\Delta m_K$ be less than about 30 \% of the SM contribution. Conservatively
assuming no near cancellations involving the terms in Eq.~(\ref{kkbar_v33}),
we obtain the following bounds:
\beq
|\theta^{(d)}_{12}| \lsim 0.01 \ ,
\label{t1212limit}
\eeq
\beq
|\theta^{(d)}_{13} \theta^{(d)}_{23}| \lsim 0.4 \times 10^{-3}
\label{t4limit}
\eeq
Values of the angles satisfying the inequalities
(\ref{t1212limit}) and (\ref{t4limit}) are plausible; these angles are
calculable, and in viable models are functions of ratios of smaller to
larger ETC scales \cite{nt,ckm}.

We note finally that early studies of $ K^0 - \bar K^0$ mixing in ETC
models, although not based on UV-complete models, often took the ETC
interactions to be vectorial. Interestingly, the studies of which we are
aware failed to observe that there would naturally be suppression of the
amplitude due to the necessity of mixing.

\subsubsection{The $\epsilon_K$ Parameter}

We turn next to the CP-violating effects, continuing to assign
$L$ and $R$ components of down-type fields to the same ETC representation.
We define the action of the CP
operator as $CP|K^0\rangle = e^{i\xi_K}|\bar K^0 \rangle$, $CP|\bar
K^0\rangle = e^{-i\xi_K}| K^0 \rangle$ on the neutral kaon, and $CP q
(CP)^{-1} = e^{i\xi_q}\gamma^0 C \bar q^T$ for a quark field $q$, where
$\xi_K$ and $\xi_q$ are convention-dependent phases (e.g. \cite{branco}).
The CP eigenstates in the neutral kaon system are given by the mixed states
$|K_{1,2}\rangle =(|K^0 \rangle \pm e^{i\xi_K} |\bar K^0 \rangle)/\sqrt{2}$,
with eigenvalues $\pm 1$ respectively.  They differ from the mass
eigenstates by small CP-violating effects (indirect CP violation), so that
the actual mass eigenstates are $|K_S \rangle = (|K_1 \rangle + \epsilon_K
|K_2\rangle)/ \sqrt{1+|\epsilon_K |^2}$ and $|K_L \rangle = (|K_2 \rangle +
\epsilon_K |K_1 \rangle)/ \sqrt{1+|\epsilon_K |^2}$ with respective masses
$m_{K_L}$ ($ = m_{K_h}$) and $m_{K_S}$ ($= m_{K_\ell}$). Making use of
approximations justified by experiment, namely $|Im(\Gamma_{12})| <<
|Im(M_{12})|$, $\Gamma_{K_S} >> \Gamma_{K_L}$, and $\Delta m_K \simeq
(1/2)\Gamma_ {K_S}$, one can derive a rephasing invariant expression for
$|\epsilon_{K}|$,
\beq
|\epsilon_K| \simeq
\frac{|{\rm Im}(M_{12}\,e^{i(\xi_K+\xi_d-\xi_s)}\,(V_{us}^{\ast}V_{ud})^2)|}
{\sqrt{2}\,\Delta m_{K}\,|V_{us}^{\ast}V_{ud}|^2}\,.
\label{eps}
\eeq
In this expression, the convention-dependent $\xi$ phases are removed by
corresponding phases in $M_{12}$. Experimentally, $|\epsilon_K| \simeq 2
\times 10^{-3}$.

The ETC contributions to ${\rm Re}(M_{12})$ are small, so that $\Delta m_K$
is determined mainly by the SM.  We use the experimental value for $\Delta m_K$ in the
denominator of Eq. (\ref{eps}). In the numerator, $M_{12}$ arises dominantly
from the SM, with ETC making a smaller contributions. We focus on the
latter.

The CKM factor $V_{us}^{\ast} V_{ud}$, defined according to
Eq.~(\ref{vckm}), enters Eq. (\ref{eps}) because it multiplies the SM
tree-level decay amplitude of $K$ mesons into a final state of two pions
with total isospin $I=0$.  We explicitly include it to denote the rephasing
invariance of $\epsilon_K$. In the canonical parametrization of the CKM
matrix, $V_{ud}$ and $V_{us}$ are real, but we use a more general form.
We expand in small angles the CKM factor (\ref{vckm}), in analogy to what we
did in Eq.~(\ref{kkbar_v33}). Keeping terms up to quadratic order, we have
\beqs
(V_{us}^{\ast} V_{ud})^2 & \simeq & (\theta_{12}^{(d)})^2 \,- \,
2 e^{-i(\beta^{(d)}_1-\beta^{(d)}_2-\beta^{(u)}_1+\beta^{(u)}_2)}\,
\theta_{12}^{(d)} \theta_{12}^{(u)} \cr\cr
& + & e^{-2i(\beta^{(d)}_1-\beta^{(d)}_2-\beta^{(u)}_1+\beta^{(u)}_2)}\,
(\theta_{12}^{(u)})^2 \ .
\eeqs

From the bound in (\ref{t1212limit}), together with the approximate relation
$|V_{us}| \sim |\theta_{12}^{(u)}-\theta_{12}^{(d)}| \simeq 0.22$, one can
infer that $|V_{us}| \simeq |\theta_{12}^{(u)}| \simeq 0.22 \gg
\theta^{(d)}_{12}$, so that, to a good approximation,
$(V_{us}^{\ast}V_{ud})$ is dominated by the term proportional to
$(\theta^{(u)}_{12})^2$, and
\beq
\frac{(V_{us}^* V_{ud})^2 }{|V_{us}^* V_{ud}|^2 }
\simeq
e^{-2i(\beta^{(d)}_1-\beta^{(d)}_2-\beta^{(u)}_1+\beta^{(u)}_2)}\,.
\eeq

While mixing angles can naturally be small in ETC theories, arising as
ratios of hierarchical ETC scales, there is no indication of a mechanism
suppressing CP-violating phases. We take them to be ${\cal O}(1)$. To
suppress the ETC contribution to $\epsilon_K$, the mixing angles must then
be smaller than required to saturate (\ref{t1212limit}) and (\ref{t4limit}),
derived from $\Delta m_K$.  From Eq.~(\ref{eps}), requiring that the ETC
contribution to $\epsilon_K$ be smaller than $30\%$ of the SM, we obtain the
bounds
\beq
|\theta^{(d)}_{12}| \lsim 10^{-3} \ ,
\label{t1212limiteps}
\eeq
\beq
|\theta^{(d)}_{13} \theta^{(d)}_{23}| \lsim  10^{-4} \ .
\label{t4limiteps}
\eeq
These constraints can be plausibly satisfied in the class of ETC models
analyzed here. We note that Eq. (\ref{t1212limiteps}) indicates that the
Cabibbo mixing angle arises dominantly from mixing in the up-quark sector.
We will discuss the direct CP-violation in the kaon system and the
associated quantity $\epsilon^{\prime}_K$ in Section~\ref{k2pi}.

The ETC mechanism for the natural suppression of flavor-changing effects
(large scales together with small mixing angles) is rather different from
the corresponding mechanism in the standard model. There, the $K^0 - \bar
K^0$ amplitude arises from box diagrams with two internal $W$ lines in the
$s$ and $t$ channels, and the small size of the imaginary part relative to
the real part is explained as a consequence of the smallness of
the charged-current couplings connecting the first- and second-generation
quarks to the third-generation quarks. As in the case of ETC contributions,
the smallness of the effect does not imply that the CP-violating phase
itself is small.

\subsection{Other Mixings}

\subsubsection{$B_d - \bar B_d$ Mixing}

The mixing amplitude $M_{12}$ in the neutral $B_d - \bar B_d$ system produces a
mass difference $\Delta m_{B_d}$, and, via its CP violating complex phase,
gives rise to CP-violation in the interference between mixing and decay
amplitudes in $B_d, \bar B_d$ decays.  When two conjugate states $B_d$ and
$\bar B_d$ decay to the same final state, the presence of state mixing between
$B_d$ and $\bar B_d$ and the resultant time-dependent oscillations produce
striking CP-asymmetries \cite{sanda}.  These have been measured at the
asymmetric $B$ factories Belle and BABAR. The cleanest mode $B_d, \bar B_d \to
J/\psi \ K_S$ yields, within the standard model, a precise measurement of the
quantity $\sin 2\beta \equiv \sin 2 \phi_1$.  These experiments are in
agreement with global SM fits.  In contrast to the situation in the neutral
kaon sector, this CP-violating effect is not small.

We analyze ETC contributions to this mixing where the $L$ and R
components of the down-type quarks transform according to the same,
fundamental representation of the ETC group. In the interaction $\bar d
\gamma_\lambda (A^\lambda)^1_3 b$, written in terms of quark mass
eigenstates, the field $(A^\lambda)^1_3$ is given by
\beqs
(A^\lambda)^1_3 & = &
e^{i(\beta_1^{(d)}-\beta_3^{(d)})} \frac{(V^\lambda)^1_3}{\sqrt{2}}
- e^{-i\delta^{(d)}}\theta_{13}^{(d)}\frac{(V^\lambda)_{d3}}{\sqrt{3}}
\cr\cr
& + & e^{i(\beta_2^{(d)}-\beta_3^{(d)})}\theta_{12}^{(d)}
\frac{(V^\lambda)^2_3}{\sqrt{2}}
- \theta_{12}^{(d)}\theta_{23}^{(d)}\ \frac{3(V^\lambda)_{d2}}{2\sqrt{6}}
\cr\cr
& + & e^{i(\beta_3^{(d)}- \beta_2^{(d)}-\delta^{(d)})}\theta_{13}^{(d)}
\theta_{23}^{(d)}\frac{(V^\lambda)^3_2}{\sqrt{2}}
\label{a13}
\eeqs
where we have kept terms involving ETC vector bosons with masses $\Lambda_3$
and $\Lambda_2$, and where we have expanded in small rotation angles.  We
focus on the $\Lambda_3$ term, which should be representative of the total
contribution.

As in $K^0 - \bar K^0$ mixing, there is no contribution to the relevant
four-fermion operators in the absence of mixing among the ETC gauge bosons.
With this effect included, the ETC contribution to $M_{12}$ is small
compared to the SM contribution. Using Eq.~(\ref{a13}), we estimate the part
of this contribution arising from the ETC mixing that generates the
off-diagional terms in the quark mass matrix. In order for it not to upset
the successful SM fit to the data (for $\Delta m_{B_d}$ and the CP-violating
asymmetry), and in light of the fact that
in the heavy-quark meson systems the  hadronic uncertainties are smaller
than in the kaon system,
we require that it be at most 0.1 compared with the measured absolute
value of $M_{12}$. This yields the constraint
\beq
|\theta_{13}^{(d)}| \lsim 1 \times 10^{-3} \ .
\label{bdmixing_limit}
\eeq
This constraint is plausibly satisfied in our ETC models for the same reason as
given above: the rotation angles are naturally small, since they are calculable
as ratios of smaller to larger ETC mass scales.  The constraint
(\ref{bdmixing_limit}) is comparable to the size of the actual CKM angle
$\theta_{13}$.

\subsubsection{$B_s - \bar B_s$ Mixing and $D^0 - \bar D^0$ Mixing}

Continuing to take the $L$ and $R$ components of the down-type quarks to
transform according to the same representation of the ETC
group, there will be no ETC contribution to $B_s - \bar B_s$ mixing without
ETC mixing. 
Experimentally, given the lower bound on $\Delta m_{B_s}$ in Eq.~(\ref{dm_values}) \cite{pdg}, ETC mixing angles are not strongly
constrained by the contributions to this process 
at the lowest scale $\Lambda_3$.
As noted in Ref. \cite{ckm}, if relatively conjugate representations are
employed for $L$ and $R$ components of down-type quarks, ETC contributions
involving the exchange of a virtual $V^3_2$ gauge boson will, even in the
absence of mixing, render $\Delta m_{B_s}$ considerably larger than the
SM prediction.

The physics of $D^0 - \bar D^0$ mixing is analogous to that of the systems
already considered, but for the replacement with up-type quarks in the
relevant operators. Again, if $L$ and $R$ components are assigned to
relatively conjugate representations, then unacceptably large contributions
arise from the exchange of ETC gauge bosons with no mixing required.  If $L$
and $R$ components of the up-type quarks are in the same (fundamental)
representation, then ETC mixing is required. Keeping mixing terms involving
the two lightest  ETC scales, and requiring that ETC contributions be
smaller than  the current upper limit (given in Eq.~(\ref{dm_values})), we
obtain
\beq
|\theta^{(u)}_{12}| \lsim  0.02 \,,
\eeq
\beq
|\theta^{(u)}_{13} \theta^{(u)}_{23}|  \lsim  10^{-3} \,.
\label{ddbar_real_bound}
\eeq
Again, the ETC models of the class being considered, with vector-like ETC
couplings, plausibly satisfy this bound.

We have now concluded that in the case of both the down-type quarks and the
up-type quarks, their assignment to vector-like ETC representations can lead to
an adequate suppression of $M^0 - \bar M^0$ mixing. (We have used the
fundamental representation, but the anti-fundamental would serve as well, as
would other vector-like representations.) It is important to stress, however,
that this cannot be done simultaneously, since it would lead to the absence of
realistic intra-family mass splittings and CKM mixing. If the up-type quarks,
for example, are assigned to a vector-like representation, say the fundamental,
then the $R$ components of the down-type quarks must be assigned to some other
representation. We have shown that the choice of the anti-fundamental would
lead to unacceptably large mixing, but one could consider other
possibilities. The key criterion is that the relevant four-fermion operators
violate the global $U(1)^3$ symmetry.  Thus, bounds such as
in~(\ref{ddbar_real_bound}) and in~(\ref{t4limiteps}) must not be considered
together. At most one of them applies -- to those quarks in a vectorial
(fundamental or anti-fundamental) ETC representation.

\subsubsection{Muonium-Antimuonium Conversion}

There is an interesting analog in the leptonic sector to these meson
systems: the muonium atom, a bound state of an anti-muon and an electron.
The phenomenology of the muonium-antimuonium  system is described by an
effective Lagrangian similar to those of the neutral meson systems. This is
the first encounter with leptons in this paper. All the processes considered
here involving leptons, including the present one, will allow the charged
leptons to be in any ETC representation. Suppose, for example, that the $L$
and $R$ components of charged leptons transform respectively
according to the fundamental
and conjugate-fundamental representations. Then the four-fermion operator
$[\bar e_L \gamma_\lambda \mu_L][\bar e_R \gamma^\lambda \mu_R]$ preserves
the $U(1)^3$ global generational symmetry inherited from ETC. It receives
contributions at scale $\Lambda_1$, with no further suppression due to
mixing. But even this makes the effect more than three orders of magnitude smaller
than the experimental upper limit on the muonium-antimuonium
amplitude~\cite{muonium}. Additional contributions to the
relevant four-fermion operators from the lowest ETC scale are suppressed by
mixing angles, and impose very mild bounds on these angles, which are easily
satisfied. If the $L$ and $R$ components are assigned to ETC representations
that do not lead to a $U(1)^3$-invariant four-fermion amplitude, then there
is no non-mixing contribution, and the mixing contributions lead to mild
bounds.

\section{Other Processes}

We next consider constraints on ETC mixing angles coming from other
processes: rare $K$ decays and leptonic transitions. In each case, the
current phenomenological constraints can be satisfied with any
ETC-representation assignment of the $L$ and $R$ components of the quarks and
charged leptons. Even when the relevant four-fermion operators induced by
ETC interactions preserve the global $U(1)^3$ symmetry, so that no ETC
mixing is required to produce them, the large scale $\Lambda_1$
suppresses them below experimental observability. Yet, in some cases
interesting constraints emerge by considering the additional contributions
which involve the lowest ETC scale $\Lambda_3$, through ETC mixing.  They
can be satisfied naturally in the class of models being considered.

\subsection{$K$ Decays}

Rare pseudoscalar meson decays have been extensively investigated
experimentally, and represent an important test of any model of new physics.
This is especially true for decays of the kaons. Other meson decays, i.e.
$B_d \rightarrow \phi K_S$, $B_s \rightarrow \mu^+ \mu^-$, do not yet
provide significant new bounds~\cite{asymmetries}.

\subsubsection{$K\rightarrow 2\pi$ and $\epsilon_K'/\epsilon_K$}
\label{k2pi}

CP violation in the neutral kaon system has conventionally been classified
as (i) indirect, occurring via the mixing of CP-even and CP-odd states to
form the mass eigenstates, manifested in the complex parameter $\epsilon_K$
defined in Eq.~(\ref{eps}), and (ii) direct, arising from the interference
between contributions containing different CP-violating phases to the decay
amplitudes of a $K$ meson into a final state with two pions, manifested in
the parameter $\epsilon_K^\prime$. Experimentally,
${\rm Re}(\epsilon_K^\prime/\epsilon_K) = (1.8 \pm 0.4) \times 10^{-3}$
\cite{pdg}.

In the standard model, direct CP violation arises at the one loop level from
penguin diagrams.  There are uncertainties in theoretical estimates of ${\rm
Re}(\epsilon_K^\prime/\epsilon_K)$ owing to difficulties in calculating the
relevant matrix elements and in choosing input values of some parameters such
as the strange quark mass \cite{bertolini}.  The experimentally measured value
of ${\rm Re}(\epsilon_K^\prime/\epsilon_K)$ is consistent with the SM
prediction, to within these one-order-of-magnitude uncertainties. Hence we can
deduce an indicative bound on ETC, by neglecting the SM contributions to
$\epsilon^{\prime}_K$, by estimating ETC contributions and by requiring that
they do not exceed the experimental result.

We define $A_{0,2}\,e^{i\delta_{0,2}}\,=\,\langle \pi \pi _{(0,2)} | H_{eff}
|K^0 \rangle$, explicitly factoring out the (CP-conserving) strong phases
$\delta_{0,2}$ due to final-state interactions. (The standard model produces
the operator $[\bar s_L\gamma_{\lambda} u_L] [\bar u_L \gamma^{\lambda} d_L]$
via tree-level exchange of a W boson.) All the relevant four-fermion operators
contain one s-quark and three first-family quarks. Thus all the operators that
arise from ETC interactions, independent of the ETC-representation assignment
of the $L$ and $R$ components, violate the $U(1)^3$ global symmetry and require
ETC mixing in order to be generated. Consider, for example, the operator $[\bar
s_L \gamma_{\lambda} d_L ] [\bar u_L \gamma^{\lambda} u_L ]$. We estimate
(using, for QCD effects, the phenomenological approach reviewed
in~\cite{bertolini}, in which the experimental result $|A_2/A_0| \simeq 0.05$
is built in) that its contribution to $A_I$ is of order $0.01 A_I^{SM}
\theta^{(d)\,L}_{13} \theta^{(d)\,L}_{23}(\theta^{(u)\,L}_{13})^2 \omega/
(V^*_{us}V_{ud})$, where $\omega$ is an ${\cal O}(1)$ phase factor containing
the phases in Eqs.~(\ref{ufchi})-(\ref{pdelta}), and where the numerical factor
$0.01$ includes the ETC scale $\Lambda_3$ as well as QCD effects.  Using the
approximate relation $|{\rm Re}(\epsilon_K^\prime/\epsilon_K)|\simeq |{\rm Im}
(A_2/A_0)|/|\sqrt{2}\epsilon_K| $, and the experimental value of
$|\epsilon_K|$, we obtain the mild constraint $
(\theta_{13}^{(d)L}\,\theta_{23}^{(d)L})^{1/2}|\theta_{13}^{(u)L}| \lsim 0.04 $
with similar bounds on other combinations of mixing angles. These bounds are
not particularly restrictive, compared with those from $\Delta m_K$ and
$\epsilon_K$.

This limit on mixing angles is milder than those
in~(\ref{t1212limiteps})-(\ref{t4limiteps}), derived from $\epsilon_K$, which
is bigger experimentally. Both quantities can be estimated as the ratio of the
CP-violating contribution to an amplitude over its CP-conserving part (the
mixing amplitude $M_{12}$ for $\epsilon_K$, the decay amplitudes $A_{0,2}$ for
$\epsilon^{\prime}_K$). The dominant CP-conserving SM contribution to $A_0$
arises at tree level, with a very modest CKM suppression, so that new physics
contributions arising at scales much larger than the electroweak are relatively
strongly suppressed. By contrast, $M_{12}$ arises in the standard model from
strongly GIM-suppressed loop diagrams, and hence is much more sensitive to new
physics at high scales, if no analog to the GIM suppression is present.

\subsubsection{$K^+ \to \pi^+ \mu^\pm e^\mp$}

Current limits are $BR(K^+ \to \pi^+ \mu^+ e^-) < 2.8 \times 10^{-11}$ and
$BR(K^+ \to \pi^+ e^+ \mu^-) < 5.2 \times 10^{-10}$ from the E865 experiment
at BNL \cite{e865_lfv}.  The $K^+ \to \pi^+ \mu^+ e^-$ decay arises from the
elementary process $\bar s \to \bar d \mu^+ e^-$ with a spectator $u$.
Because both the $L$ and $R$ components of the down-type quarks and leptons
enter the amplitudes for this process, there will typically be an amplitude
invariant under the $U(1)^3$ generational symmetry. Only in the case that
the $L$ and $R$ components of the quarks are assigned to the same
ETC-representation, and the $L$ and $R$ components of the charged leptons are
also assigned to the same representation (but conjugate to that of the
down-type quarks), will this not be the case. Excluding this possibility to
focus on the worst case, the amplitude can occur with no ETC mixing, and is
of order $1/\Lambda_1^2$. But even this is sufficiently small relative to
the above experimental limit. Since this contribution, without ETC mixing,
is well below the experimental bound, one can anticipate that contributions
involving mixing will not lead to especially tight constraints on the mixing
angles. An estimate as in Section IV, expanding in small rotation angles,
leads to a bound much milder than those derived so far.

The situation with $K^+ \to \pi^+ \mu^- e^+$ is much the same.  Except for
one possible choice of ETC-representation assignments, there will be a
$U(1)^3$-invariant amplitude, with no ETC mixing required, and with
coefficient $1/\Lambda_1^2$. Again, this contribution is below the
experimental bound, and the additional contribution involving mixing leads
to only a mild bound on mixing angles.

\subsubsection{$K_L \to \mu^\pm e^\mp$}

The current limit on the branching ratio is $BR(K_L \to \mu^\pm e^\mp) <
4.7 \times 10^{-12}$ from the E871 experiment at BNL \cite{e871_lfv}. The
peculiarity of this process is that the hadronic matrix element arises only
from the axial-vector part of the relevant bilinear quark operator.

In the case where ETC interactions of the quarks are not vectorial, there is
always a contribution to these decays at scale $\Lambda_1$, with no mixing
required, because it involves a second generation fermion (or antifermion)
in both the
initial and final state. Thus one of the possible four-fermion operators
preserves the $U(1)^3$ global symmetry. Given the value of $\Lambda_1$ in
Eq.~(\ref{lamscales}), we estimate this to lead to a branching ratio $BR(K_L
\to \mu^\pm e^\mp) \approx 10^{-12}$, still allowed by current experimental
bounds, but potentially observable in next-generation experiments. If the
ETC coupling to the down-type quarks is vectorial, this process cannot be
generated to any order in ETC interactions, even if the relevant
four-fermion operator does not violate $U(1)^3$. Nonzero contributions
depending on ETC interactions arise from graphs involving both electroweak
gauge bosons and ETC exchange. These are expected to yield an amplitude of
order $(\alpha/\pi)(1/\Lambda_1^2)$ and hence a rate that is safely smaller
than the above experimental limit.

\subsubsection{$K^+ \to \pi^+ \nu\bar\nu$}

Here we consider the decay $K^+ \to \pi^+ +$ missing (weakly interacting)
neutrals.  The branching ratio for this decay has been measured by the E787 and
E949 experiments at BNL, with the result \cite{e787} $BR(K^+ \to \pi^+ + {\rm
missing \ neutrals}) = (1.57^{+1.75}_{-0.82}) \times 10^{-10}$.  In the
standard model the rate for this observed decay is the (incoherent) sum of the
rates for each of the three individual decays $K^+ \to \pi^+ \nu_\ell
\bar\nu_\ell$, where $\nu_\ell$ are the three light
neutrinos.  These
decays arise at the one-loop level.  The SM prediction for the branching ratio
is $\sum_\ell BR(K^+ \to \pi^+ \nu_\ell \bar\nu_\ell) = (0.77 \pm 0.11) \times
10^{-10}$ \cite{gi}, consistent with the measurement. 
In this process and in others involving neutrino final
states, we neglect neutrino masses, 
since they are very small compared to the other masses in the process.

Whether the $L$ and $R$ components of the down-type quarks are assigned to the
same or conjugate ETC representations, there will be a contribution to the
underlying process $\bar s \rightarrow \bar d + \nu_{e} \bar\nu_{\mu}$ with
no ETC mixing and proportional to $1/\Lambda_1^2$ . It can proceed, for
example, by the exchange of a single $V^1_2$ ETC gauge boson. With
$\Lambda_1 \approx 10^{3}$ TeV, the ETC contribution is comfortably below
the experimental bound.

Other contributions arise from ETC mixing, for example those generating
off-diagonal quark mass terms. The process can proceed by $\bar s$ emitting
a virtual $V_{d3}$ ETC gauge boson, going to a $\bar d$. The $V_{d3}$ can
then produce, with no further ETC mixing required, the pair of interaction
eigenstates $\nu_\tau \bar \nu_\tau$. Other contributions include the
exchange of a $V_{d2}$ ETC gauge boson, but do not 
impose significant bounds. Higher-order contributions are understood to be
present, owing to the strong-coupling nature of the ETC theory, but these
have the same overall generational index structure as the lowest-order
exchanges. Requiring  the contribution induced by  $V_{d3}$ exchange 
 be less than the SM prediction for the branching ratio,
we obtain the limit
\beq |\theta_{13}^{(d)\chi} \theta_{23}^{(d)\chi^{\prime}}| \lsim 10^{-3} \,,
 \label{kpnnlimit} \eeq
where we retain the labels $\chi,\chi'=L,R$,
because the bound does not depend on the chirality assignments.

These bounds are somewhat weaker than the bound we derived earlier from the
measurement of $\epsilon_K$ in Eq. (\ref{t4limiteps}).  Since plausible
values of the above angles suggest that the left-hand side of eq.
(\ref{kpnnlimit}) could nearly saturate the limit, it is possible that ETC
contributions to this decay could amount to a significant fraction of the SM
branching ratio.

\subsubsection{$K_L\rightarrow \pi^0 \nu \bar \nu$}

This is, experimentally, the decay $K_L \to \pi^0 +$ missing weakly
interacting neutrals. In the standard model, it is $K_L \to \sum_\ell \pi^0
\nu_\ell \bar\nu_\ell$.  This is of interest because, although it is not
manifestly a CP-violating decay, the main contribution in the standard model
turns out to involve a direct CP-violating amplitude \cite{kopio}.  As with
$K^+ \to \pi^+ \nu\bar\nu$, the amplitude can be calculated accurately in
the standard model terms of the CKM mixing parameters. From a current global
fit, one obtains the SM prediction $BR(K_L \to \sum_\ell \pi^0 \nu_\ell
\bar\nu_\ell) = (2.6 \pm 0.5) \times 10^{-12}$ \cite{gi}.  The current
experimental limit, $BR(K_L \to \sum_\ell \pi^0 \nu_\ell \bar\nu_\ell) < 5.9
\times 10^{-7}$ \cite{pdg}, is not nearly sensitive to the SM prediction,
but the future KOPIO experiment at BNL plans to measure the decay and test
this prediction. The ETC contribution to the decay $K_L \to \pi^0 +$ missing
weakly interacting neutrals arise in a manner similar to that for $K^+ \to
\pi^+ +$ missing neutrals and might ultimately produce a measurable
deviation relative to the SM prediction.

\subsection{Leptonic Processes}

ETC-boson exchanges analogous to those discussed for quarks induce also
four-fermion operators of relevance for experimentally accessible leptonic
processes. They lead to only mild constraints, and, as with the semi-leptonic
processes above, allow the $L$ and $R$ components of the charged leptons to be
in any representation of the ETC group.

\subsubsection{$\mu^+ \to \lowercase{e}^+ \lowercase{e}^+
\lowercase{e}^-$}

Experimentally, $BR(\mu^+ \to e^+ e^+ e^-) < 1.0 \times 10^{-12}$. Since
this process involves only a single second-generation fermion, the
four-fermion amplitudes contributing to it necessarily violate the $U(1)^3$
global symmetry. Thus ETC mixing is required whatever the ETC-representation
assignments of the $L$ and $R$ components of the charged leptons. Proceeding as
in Section IV, keeping terms involving the exchange of ETC vector bosons of
the lowest two masses, $\Lambda_3$ and $\Lambda_2$, and performing an
expansion in small rotation angles, we have the following ETC contribution
to the effective Hamiltonian for $\mu^- \to e^- e^- e^+$:
\beq\biggl [
\frac{16}{3\Lambda_3^2}e^{-i\delta^{(\ell)}} (\theta_{13}^{(\ell)})^3
\theta_{23}^{(\ell)}   + 
\frac{6}{\Lambda_2^2}(\theta^{(\ell)}_{12})^3 \biggr ][\bar e \gamma_\lambda
\mu][\bar e \gamma^\lambda e]  \ ,\label{mu3eamp} \eeq
where we have dropped chirality labels on the mixing angles 
for simplicity. From the experimental upper limit
on $\mu^+ \to e^+ e^+ e^-$, we get the bound
$|\theta_{13}^{(\ell)}|^{3/2}|\theta_{23}^{(\ell)}|^{1/2} \lsim 0.006$. This
is a relatively mild bound, and easily accommodated in our models where
mixing angles are ratios of hierarchical ETC scales.

\subsubsection{$\mu \to \lowercase{e}$ Conversion}

A number of searches have been carried out for $\mu \to e$ conversion in the
Coulomb field of a nucleus.  One contribution to $\mu \to e$ conversion is
from a process in which $\mu \to e$ plus a virtual photon, which is
exchanged with the nucleus.  In ETC theories this process arises from the
same type of amplitude that produces $\mu \to e \gamma$, which we have
bounded earlier in Ref. \cite{dml}. We focus here on an additional ETC
contribution in which, via mixing, the $\mu$ makes a transition to an $e$
with the exchange of a virtual $V_{d3}$ with the nucleons. This process,
involving only a single second-generation fermion, violates the $U(1)^3$
symmetry, and hence requires ETC-mixing to be generated.

One can write the effective Hamiltonian for the $\mu \to e$ conversion
process as
\beq
{\cal H}_{\mu e} = \frac{G_F}{\sqrt{2}}  \sum_{i=0,1} \biggl [
\bar e \gamma_\lambda(g_{VV}^{(i)}-g_{AV}^{(i)}\gamma_5)\mu
J_{V,nuc.}^\lambda
+
\bar e \gamma_\lambda(g_{VA}^{(i)}-g_{AA}^{(i)}\gamma_5)\mu
J_{A,nuc.}^\lambda
\biggr ]
\label{hme}
\eeq
where $i=0,1$ refer to isoscalar and isovector contributions and
$J_{V,nucl.}$ and $J_{A,nucl.}$ denote the effective nuclear vector and
axial vector currents.  Experimental bounds imply limits such as
$g_{VV}^{(0)} < 4 \times 10^{-7}$ \cite{sindrum}.  From these we derive the
bound $[\theta^{(\ell)}_{13} \theta^{(\ell)}_{23}]^{1/2} |\theta^{(f)}_{13}|
\lsim 0.006$, comparable to that from $\mu^+ \to e^+ e^+ e^-$, with the
difference that it involves both charged lepton and quark mixing angles. We
have again dropped chirality labels on the mixing angles for simplicity.
The future MECO experiment at BNL projects a large improvement in
sensitivity in the search for $\mu \to e$ conversion \cite{meco}.

\subsubsection{Ordinary $\mu$ Decay}

The exchange of virtual ETC gauge bosons adds new contributions to ordinary
$\mu$ decay, thus modifying the effective Fermi coupling with respect to the
standard model. Although $\mu^+$ decay is conventionally regarded as being
$\mu^+ \to e^+ \bar \nu_\mu \nu_e$, as predicted by the standard model, at an
experimental level it is simply $\mu^+ \to e^+ +$ unobserved weakly interacting
neutrals.  On the other hand, since the corrections to the SM decay rate we are
considering are very small, the most important ETC contributions are the
coherent ones (to the same final states as the standard model), so that we
focus our attention on the operator $[\bar \mu_L \gamma_\lambda e_L] [\bar
\nu_{e L} \gamma^\lambda \nu_{\mu L}]$, which receives contributions both from
$W$ boson exchange and from ETC exchange.

This is another process in which ETC can contribute without any mixing
required, since the relevant operator preserves the $U(1)^3$ global
symmetry. The SM coefficient is $G_F/\sqrt{2} = g^2/(8 m_W^2)$. ETC exchange at
scale $\Lambda_1$ gives a contribution $\simeq 8/\Lambda_1^2 \simeq 10^{-6}
G_F/\sqrt{2}$, and hence is negligible. There are also terms due to lepton
mixing arising from much lower scales, such as the one involving the exchange
of a virtual $V_{d3}$. In order not to modify substantially electroweak
precision observables, we require ETC to contribute at most 0.1 \% of the SM
amplitude, and derive a weak bound compared to those of previous subsections,
which is easily satisfied in our ETC models,

\section{Discussion and Conclusions}

Constraints from neutral flavor-changing processes were among the first
concerns in studies of extended technicolor. In this paper we have
reconsidered these constraints, focusing on the relevant four-fermion
operators. We have taken account of the multi-scale nature of the ETC gauge
symmetry breaking, in a class of ultra-violet complete models in which the
TC theory is approximately conformal ("walking"), and the ETC gauge group
commutes with the SM gauge group.
Features such as intra-family mass splitting and
CKM mixing are generated by
the dependence of the ETC representation assignments
of the SM fermions on their assignments
under the standard model. This work extends our earlier general
study in Ref. \cite{ckm} and specific studies of (dimension-five) dipole
moment operators in Refs. \cite{dml} and \cite{qdml}.

We have described an approximate global generational $U(1)^3$ symmetry,
inherited by the low-energy effective theory from the underlying ETC gauge
symmetry, that controls the coefficients of the four-fermion operators.
Operators that violate this symmetry are suppressed not only by large ETC
scales, but also by (small) mixing effects among ETC gauge bosons. Employing
this symmetry classification, we have considered two, relatively simple types
of assignments: those in which $L$ and $R$ components of quark (and
techniquark) fields of a given charge transform according to the same
(fundamental or anti-fundamental) representation of the ETC group, and those in
which they transform according to the opposite (fundamental and
anti-fundamental) ETC representations. Corresponding assignments for the
charged leptons must also be made, but the choice is not critical for the
phenomenology of this paper.

We have analyzed $K^0 - \bar K^0$ and $B_d - \bar B_d$ mixings, and limits on
$B_s - \bar B_s$ and $D^0 - \bar D^0$ mixing. We have also considered the
decays $K^+ \to \pi^+ \mu^\pm e^\mp$, $K_L \to \mu^\pm e^\mp$, $\mu^+ \to e^+
e^+ e^-$, $\mu \to e$ conversion in the field of a nucleus, and effects on
$\mu$ decay. ETC contributions are suppressed by the heaviness of the ETC
scales involved, and for operators that violate the $U(1)^3$ by the smallness
of the requisite ETC mixing effects.

For the case in which $L$ and $R$ components of quarks of a given electric
charge transform according to relatively conjugate representations, some
dangerous four-fermion operators involving SM fields preserve the global
$U(1)^3$ symmetry, and hence are suppressed only by the ETC scales, with no
mixing required. For example, in the case of down-type quarks this leads to a
very large contribution to the $K^0 - \bar K^0$ mixing amplitude, excluding the
viability of such an assignment for $s$ and $d$ quarks. We note, however, that
this assignment for the down-type quarks, with up-type quarks coupling
vectorially to ETC, produces charged-current (CKM) flavor mixing, together with
substantial intra-generational mass splitting such as $m_t \gg m_b$. The latter
is achieved without having introduced new sources of custodial-SU(2) violation
below the (large) ETC scales, and hence suppressing new physics contributions
to the $\rho$ parameter in precision electroweak physics.

For the other case, in which $L$ and $R$ components transform according to the
same (fundamental or anti-fundamental) representation, no excessively large
contributions to any of the processes we have considered are generated provided
the ETC scales are large enough and ETC mixing effects are small enough.  This
is because the global $U(1)^3$ symmetry forbids the most dangerous four-fermion
operators, which can then arise only through ETC-mixing. This fact seems not to
have been noticed in earlier studies of ETC theories. We have focused on those
ETC mixings leading to the off-diagonal structure of the quark mass matrices,
i.e. on the mixing angles parameterizing the unitary matrices that diagonalize
these matrices. Interesting bounds on these mixing angles emerge in the up- and
down- sectors separately, and thus constrain even mixing parameters that do not
enter in the CKM matrix.

But we stress that one cannot simultaneously assign both the down-type quarks
and the up-type quarks, to vector-like ETC representations, since it would,
without additional ingredients, lead to the absence of realistic intra-family
mass splittings and CKM mixing. If the up-type quarks, for example, are
assigned to a vector-like representation, say the fundamental, then the $R$
components of the down-type quarks must be assigned to some other
representation. We have shown that the choice of the anti-fundamental would
lead to unacceptably large $M^0 - \bar M^0$ mixing, but one could consider
other possibilities. The key criteria are that the ETC scales be large enough
while still generating realistic fermion masses (as in the present paper), {\it
and} that the four-fermion operators describing $M^0 - \bar M^0$ mixing violate
the global $U(1)^3$ symmetry. Thus, bounds such as in~(\ref{ddbar_real_bound})
and in~(\ref{t4limiteps}) must not be considered together. At most one of them
applies -- to those quarks in a vectorial (fundamental or anti-fundamental) ETC
representation. The bounds on their mixing angles constrain future ETC
model-building, but can naturally be satisfied in the class of models
considered here. Furthermore, values of fermion mixing angles consistent with
our constraints still allow substantial deviations from the SM predictions, for
both CP-conserving quantities such as $\Delta m_K$, $K^+ \to \pi^+ \mu^+ e^-$,
$K^+ \to \sum_\ell \pi^+ \nu\bar\nu$, and similar mixings and decays, and for
CP-violating quantities.

This suggests a direction for future ETC model building within the class
considered here. One should try, in effect, to capture the best features of the
above representation choices. The $L$ and $R$ components of quarks should be
assigned to representations of the ETC gauge group in such a way as to suppress
the down-type quark mass matrix relative to the up-type mass matrix, at least
those parts that determine the b-quark mass relative to the t-quark. The
assignments must also suppress dangerous four-fermion operators, specifically
those describing $M^0 - \bar M^0$ mixing, that is, they must render the
relevant four-fermion operators non-invariant under the $U(1)^3$ symmetry.  We
are currently exploring this possibility.

Ultraviolet-complete ETC theories take on a very ambitious task, to explain
dynamically, with a very small number of parameters, electroweak
symmetry breaking, fermion generations, masses, and mixing. It is not
surprising that it is difficult to find an entirely successful model, but we
believe that there is strong motivation to continue the search. The
constraints that we have obtained in our previous papers and in the present
analysis may provide some helpful guidance for this model-building
enterprise.

\acknowledgements

We thank K.~Lane and Yang Bai for stimulating discussions and acknowledge the
Aspen Center for Physics, where some of this work was done.  This research was
partially supported by the grants DE-FG02-92ER-4074 (T.A., M.P.) and
NSF-PHY-00-98527 (R.S.).

\vfill
\eject


\begin{thebibliography}{99}

\bibitem{etc}
S. Dimopoulos, L. Susskind, Nucl. Phys. {\bf B155}, 23, (1979);
E. Eichten, K. Lane, Phys.  Lett. B {\bf 90}, 125 (1980).

\bibitem{wtc}
B. Holdom, Phys. Lett. B {\bf 150}, 301 (1985); K Yamawaki,
M. Bando, K. Matumoto, Phys. Rev. Lett. {\bf 56}, 1335 (1986);
T. Appelquist, D. Karabali, L.C.R. Wijewardhana, Phys. Rev. Lett. {\bf
57}, 957 (1986); T. Appelquist and L.C.R. Wijewardhana, Phys. Rev. D
{\bf 35},
774 (1987); Phys. Rev. D {\bf 36}, 568 (1987).

\bibitem{nf} Our use of SU(2)$_{TC}$ with one SM family of technifermions
($N_f=8$ technidoublets) is based on the inference from the two-loop beta
function and gap equations that this theory confines, spontaneously breaks
chiral symmetry, and has an approximate infrared-stable fixed point that gives
rise to walking behavior.  Efforts continue to try to use lattice gauge theory
methods to answer the question, in a (vectorial) SU($N$) gauge theory, of what
the critical value is for the number of massless fermions in the fundamental
representation at which the theory goes over from a phase with confinement and
spontaneous chiral symmetry breaking to a nonabelian Coulomb phase; see, e.g.,
R.  Mawhinney, Nucl. Phys. Proc.  Suppl. {\bf 83}, 57 (2000) and Y.  Iwasaki et
al., Phys. Rev. D {\bf 69}, 014507 (2004).

\bibitem{precision}
T. Appelquist and J. Terning, Phys. Lett. {\bf B315}, 139
(1993); T. Appelquist, J. Terning, L.C.R. Wijewardhana, Phys. Rev.
Lett.  {\bf
77}, 1214 (1996); {\it ibid.} {\bf 79}, 2767 (1997).

\bibitem{at94}
T. Appelquist, J. Terning, Phys. Rev. D {\bf 50}, 2116 (1994).

\bibitem{nt}
T. Appelquist and R. Shrock, Phys. Lett. {\bf B548}, 204 (2002).

\bibitem{lrs}
T. Appelquist and R. Shrock, Phys. Rev. Lett. {\bf 90}, 201801 (2003).

\bibitem{ckm}
T.~Appelquist, M.~Piai and R.~Shrock,
Phys.\ Rev.\ D {\bf 69}, 015002 (2004)
[hep-ph/0308061].

\bibitem{dml}
T.~Appelquist, M.~Piai and R.~Shrock,
Phys.\ Lett.\ B {\bf 593}, 175 (2004)
[hep-ph/0401114].

\bibitem{qdml}
T.~Appelquist, M.~Piai and R.~Shrock,
Phys.\ Lett.\ B {\bf 595}, 442 (2004)
[hep-ph/0406032].

\bibitem{ssvz}
P.~Sikivie, L.~Susskind, M.~B.~Voloshin and V.~I.~Zakharov,
Nucl.\ Phys.\ B {\bf 173}, 189 (1980).

\bibitem{pdg}
See http://pdg.lbl.gov.

\bibitem{lane_kkb}
For recent discussions of CP violation in the ETC context, see K. Lane,
hep-ph/0202255; A. Martin and K. Lane, hep-ph/0404107.


\bibitem{parodi} F. Parodi, in {\it Proc. Fifth International Conference on
Hyperons, Charm, and Beauty Hadrons}, ed. C. Kalman et al., Nucl. Phys.  B
(Proc. Suppl.)  {\bf 115} (2003) 212; M. Battaglia et al., hep-ph/0304132;
M. Ciuchini et al., hep-ph/0307195; http://cern.ch/ckm-workshop; M. Bona et
al., hep-ph/0408079; Y. Sakai (Belle) and M. Giorgi (BABAR), talks at the XXVI
Int. Conf. on High Energy Physics, Beijing, available at
http://ichep04.ihep.ac.cn.  


\bibitem{pitts}
We use the updated values for $B_d$ and $B_s$ mixing from the
Heavy Flavor Averaging Group, http://www.slac.stanford.edu/xorg/hfag/ which
have been incorporated in the 2004 PDG \cite{pdg}.

\bibitem{st}
R. Shrock and S. Treiman, Phys. Rev. D {\bf 19}, 2148 (1979).

\bibitem{lattice}
For recent reviews, see N. Yamada, in {\it Lattice 2002 -
Proceedings of the XXth International Symposium on Lattice Field Theory},
eds. R. Edwards, et al., Nucl. Phys. B (Proc. Suppl.) {\bf 119}, 93 (2003);
D. Becirevic in {\it Lattice 2003 - Proceedings of the XXI International
Symposium on Lattice Field Theory}, ed. S. Aoki et al., Nucl. Phys. B
Proc. Suppl. {\bf 129}-{\bf 130}, 34 (2004).

\bibitem{branco} See, e.g., G. Branco, L. Lavoura, and J. Silva, {\it CP
Violation} (Oxford Univ. Press, Oxford, 1999); I. Bigi and A. I. Sanda, {\it CP
Violation} (Cambridge Univ. Press, Cambridge, 2000).

\bibitem{sanda}
A. Carter and A. I. Sanda, Phys. Rev. Lett. {\bf 45}, 952 (1980);
Phys. Rev. D {\bf 23}, 1567 (1981).

\bibitem{muonium}
Y.~Kuno and Y.~Okada,
Rev.\ Mod.\ Phys.\  {\bf 73}, 151 (2001) [hep-ph/9909265];
L.~Willmann {\it et al.},
Phys.\ Rev.\ Lett.\  {\bf 82}, 49 (1999) [hep-ex/9807011].


\bibitem{asymmetries}
Determinations of $\sin 2\beta$ from $b \to s c
\bar c$ decay modes such as $B_d^0, \ \bar B_d^0 \to J/\psi \ K_S$ yield $\sin
2\beta = 0.726 \pm 0.037$ while determinations from $b \to s q \bar q$ penguin
modes, such as $B_d \to \phi K_S$, $B_d \to (K \bar K)_{nonres.}K_S$ and $B_d
\to \eta' K_S$, yield $\sin 2 \beta = 0.42 \pm 0.10$ (BABAR) and $\sin 2\beta =
0.43^{+12}_{-0.11}$ (Belle).  The latter values differ by about $2.5\sigma$
from the $b \to s c \bar c$ value of $\sin 2 \beta$.
New physics affecting the decay amplitudes
of these processes, in particular in the class of ETC theories we 
are discussing~\cite{qdml}, could play a role in explaining this difference.

\bibitem{bertolini}
See, e.g., S.~Bertolini, M.~Fabbrichesi and J.~O.~Eeg,
Rev. Mod. Phys.  {\bf 72}, 65 (2000).

\bibitem{e865_lfv}
R. Appel et al., Phys. Rev. Lett. {\bf 85}, 2450 (2000).
Analysis of further BNL
E865 data reduces slightly the upper limit on $K^+\rightarrow
\pi^+ \mu^+ e^-$ (M.~Zeller,
private communication; A.~Sher thesis;
E865 Collab. to be published.)

\bibitem{e871_lfv}
D. Ambrose et al., Phys. Rev. Lett. {\bf 81}, 5734 (1998).

\bibitem{e787}
S. Adler et al., Phys. Rev. Lett. {\bf 79}, 2204 (1997); {\it ibid}
{\bf 84}, 3768 (2000); {\it ibid} {\bf 88}, 041803 (2002);
hep-ex/0403034; V. Anisimovsky et al., hep-ex/0403036.

\bibitem{gi}
See, e.g., G. Isidori, hep-ph/0307014.

\bibitem{kopio}
L. Littenberg, Phys. Rev. D {\bf 39}, 3322 (1989); hep-ex/0201026,
hep-ex/0212005; \goodbreak http://www.bnl.gov/rsvp/KOPIO.htm.

\bibitem{sindrum}
C. Dohmen et al., Phys. Lett. B {\bf 317}, 631 (1993); W. Honecker et
al.,
Phys. Rev. Lett. {\bf 76}, 200 (1996); A. van der Schaaf, Nucl. Instr.
Meth.
A {\bf 503}, 281 (2003).

\bibitem{meco}
See http://meco.ps.uci.edu.

\end{thebibliography}
\end{document}